# SAT problem and statistical distance

FENG PAN Donghua University, Shanghai, China

**Abstract**: In this paper with two equivalent representations of the information contained by a SAT formula, the reason why string generated by succinct SAT formula can be greatly compressed is firstly presented based on Kolmogorov complexity theory. Then what strings can be greatly compressed were classified and discussed. In this way we discovered the SAT problem was composed of a basic problem: distinguish between two different distributions induced under the computer. We then tried to map this problem into quantum mechanics, or the quantum version of the basic problem: this time two different distributions are induced under quantum mechanics. Based on the equivalence of statistical distance between probability space and Hilbert space, in the same time this distance is invariant under all unitary transformations. The basic problem's quantum version cannot be efficiently solved by any quantum computer. In the worst case, any quantum computer must perform exponential times measurement in order to solve it. In the end we proposed the main theorem : The statistical distance in SAT formula space and probability space are identical. We tried to prove it using the relationship of Kolmogorov complexity and entropy. It showed there is no difference to solve the basic problem in SAT formula space or probability space. In the worst case, exponential trials must be performed to solve it. NP!=P.

## 1. INTRODUCTION

As we know, any function $f:\{0,1\}^n \to \{0,1\}$ can be constructed from the elementary gates AND, OR, NOT and FANOUT [1]. Therefore, these constitute a universal set of gates for classical computation. There are $2^{2^n}$ possible functions in this function space.

By representing computation using circuits, it's easy to show that some functions require very large circuit to compute [2].

**Theorem 1**: For every $n$>1, there exists a function $f:\{0,1\}^n \to \{0,1\}$ that cannot be computed by a circuit $C$ of size $2^n/(10n)$.

There is another way to phrase it and yield a stronger result than the Theorem 1: not only does there exist a hard function (not computed by $2^n/(10n)$ size circuits), but in fact the vast majority of functions form $\{0,1\}^n$ to $\{0,1\}$ are hard [1].

The Boolean Satisfiability Problem (abbreviated as SAT) is the problem of determining if there exists an interpretation that satisfies a given Boolean formula. In other words, it asks whether the variables of a given Boolean formula can be consistently replaced by the values TRUE or FALSE in such a way that the formula evaluates to TRUE. If this is the case, the formula is called satisfiable. On the other hand, if no such assignment exists, the function expressed by the formula is identically FALSE for all possible variable assignments and the formula is unsatisfiable. SAT is one of the first problems that were proven to be **NP**-complete. And resolving the question whether SAT has an efficient algorithm is equivalent to the **P** versus **NP** problem [3]. As we can see, the SAT problem wants to know the formula, with the $2^n$ input, is or is not the one output the $2^n$ zeros (represents unsatisfiable). The SAT, in this regard, wants to know the information of the whole function which corresponds to $2^n$ inputs.

Instead of modeling Boolean circuits as labled graphs, we can also model them as a straight-line program [1]. A program is straight-line if it contains no branching or loop operations (such as “if” or “goto”), and hence its running time is bounded by the number of instructions it contains. The equivalence between Boolean circuits and straight-line programs is fairly general and holds for essentially any reasonable programming language. The straight-line program can be obviously demonstrated with Boolean operations(**OP**). A Boolean straight-line program of length *T* with input variable $y_1, y_2, ..., y_n \in \{0,1\}$ is a sequence of *T* statements of with **OP**, where **OP** is either AND, OR, NOT and FANOUT. That’s to say, a SAT formula. So SAT formula and circuit are two equivalent ways to represent a function.

Kolmogorov defined the algorithmic (descriptive) complexity of an object to be the length of the shortest binary computer program that describes the object [4]. It’s the intrinsic descriptive complexity of an object. To be specific, The Kolmogorov complexity $K_U(x)$ of a string *x* with respect to a universal computer *U* is defined as the minimum length over all programs that print *x* and halt.

$$K_U(x) = \min_{p:U(p)=x} l(p) \tag{2}$$

Thus, $K_U(x)$ is the shortest description length of *x* over all descriptions interpreted by computer *U*. The expected length of the shortest binary computer description of a random variable is approximately equal to its entropy.

There is a profound related law of large numbers for the Kolmogorov complexity of Bernoulli $(\gamma)$ sequences. The Kolmogorov complexity of a sequence of binary random variables drawn i.i.d. according to a Bernoulli $(\gamma)$ process is close to the entropy $H_0(\gamma)$.

**Theorem 2**: (Relationship of Kolmogorov complexity and entropy)Let $x_1, x_2, ..., x_{2^n}$ be drawn according to a Bernoulli $(\gamma)$ process. Then

$$\frac{1}{2^n} K(x_1, x_2, ..., x_{2^n} | 2^n) \to H_0(\gamma) \qquad ()$$

This notion of intrinsic complexity is computer independent. Besides that, The Kolmogorov theory is very beautiful and has produced many profound and useful results that wait for further mining. Kolmogorov complexity often gives us a framework that helps us understand not only computation but efficient computation and in the same time allows us to put quite different and complex concepts in a common framework.

In all the $2^{2^n}$ possible functions, we already know the vast majority are hard and have exponential complexity measured by circuit. But in practice we are only or more interested in those functions which can be implemented by succinct circuits, then what properties those functions have? What makes them special? This question should be important and has both theory and practical value. We will first try to answer this question by using Kolmogorov theory as a bridge.

This paper is organized as the follows. In Sec.2 we focus on the reason why the string generated by succinct SAT formula can be greatly compressed. In Sec.3 we concentrate on what strings can be greatly compressed. In Sec.4 we showed the SAT problem is composed of a basic problem. In Sec.5 the quantum version of the basic problem was probed. In Sec.6 the main theorem was proposed. Sec.7 is devoted to summary and discussion.

## 2. TWO EQUIVALENT REPRESENTATIONS OF THE INFORMATION CONTAINED IN A SAT FORMULA

With the Kolmogorov complexity theory, the information contained in a specific SAT formula (function) can be represented by the following two equivalent ways. One is in the form of program (Program 1).

```
for i= 0 to 2^n − 1
    With i as the input, compute the one bit output of a SAT formula;
     Print the output bit;
end
```

This program will print out this function's corresponding $2^n$ output bit string. The only two variables in the program are *i* and the specific SAT formula. The total length of this program is:

$$l(p) = c + log\, 2^n + l(\text{SAT formula}) = c + n + O(P(n)) = O(P(n)) \qquad (3)$$

In which *c* is constant, $l(\text{SAT formula})$ represents the length of SAT formula. As we know, we're only interested in those succinct formulas, that's to say, their lengths are with polynomial(P) complexity, let us represent its length with $O(P(n))$ bit. It should be noted that the length of SAT formula should not be the direct length of that formula, if the formula is only seemingly exponential long but it can be generated by succinct program, then it is still succinct and short. It's same for circuit representation, some circuit may have exponential size, but they have a succinct representation in terms of a Tuning Machine[1], which can systematically generate any required vertex of the circuit in polynomial time.

Now we have a program which can generate the corresponding string, but according to Kolmogorov complexity theory, we can never guarantee this program is the shortest one which can generate the same $2^n$ bit string. So we have the upper bound of its Kolmogorov complexity:

$$K(x_1, x_2, ..., x_{2^n} | 2^n) \le O(P(n)) \tag{4}$$

In the same time, Kolmogorov complexity theory told us that the probability that a string can be compressed by more than *k* bits is no greater than $2^{-k}$.

**Theorem 3**: Let $x_1, x_2, ..., x_{2^n}$ be drawn according to a Bernoulli $(\frac{1}{2})$ process. Then

$$(K(x_1, x_2, ..., x_{2^n} | 2^n) < 2^n - k) < 2^{-k} \tag{5}$$

The second form that can represent the information contained in a specific SAT formula is its $2^n$ corresponding output string. According to Theorem 2, in all the possible $2^{2^n}$ output strings, the possibility of those strings that can be compressed over *k* bits will not exceed $2^{-k}$.

Then similar conclusion with Theorem 1 can be derived directly: of all the possible $2^{2^n}$ functions (strings), the length of vast majority corresponding SAT formulas are exponential long (hard), because the strings that can be greatly compressed (compressed from exponential long to **P** complexity long) are very rare.

In the same time, besides the above conclusion which Theorem 1 has already told us, we can get more useful conclusion. The SAT formula with **P** complexity cannot generate all the $2^{2^n}$ functions. It can only generate at most $2^{O(P(n))}$ different functions. Considering the fact that some formula can also be compressed, the equivalent functions generated will be less than that number. And the functions generated must have corresponding output strings that can be greatly compressed. These strings can be compressed by at least $2^n - O(P(n))$ bit and have at most the possibility of $2^{O(P(n))\text{-}2^n}$. With a view to the fact that $O(P(n)) << 2^n \quad when\ n \to \infty$, we call it greatly compressed.

## 3. THE STRINGS THAT CAN BE GREATLY COMPRESSED

Since we are more interested in succinct SAT formulas whose corresponding strings can be greatly compressed, let's look what strings can be greatly compressed. As we already know, they are very rare and special. We classify them into two types based on the number of ones (or zeros) in the sequences. Suppose the string $x = \{x_1, x_2, ..., x_{2^n}\}$ and $\sum_{i=1}^{2^n} x_i = k$ , there are $k$ ones in the string.

**Type 1** are those strings with $k = O(P(n)), \quad (k \ll 2^{n-1})$ , they can be greatly compressed according to Kolmogorov theory.

**Type 2** are those strings with $k \approx 2^{n-1}$ and meanwhile can be generated by short programs like Program 1.

As we noted, the vast majority of $2^{2^n}$ possible functions have $k \approx 2^{n-1}$. As for the type 2 strings, there are also numerous strings which can be generated by short programs, eg, $\pi, e, \sqrt{2}$ and etc, but we don't know is whether these strings can be generated by programs in the form similar to Program 1 which has $2^n$ cycles. What we do know is that there exist type 2 strings. For example, the following strings can be generated by succinct SAT formula.

0101010101010101010101010101010101010101…

0011001100110011001100110011001100110011…

Suppose $n = 2m$, if we perform the ">" operation between the first $m$ bit and the second $m$ bit, The function fulfill the problem of compare operation (>,!=,>=,…) can also be represented by succinct SAT formula.

However, in this paper we'll lay stress on type 1 and neglect type 2 strings for two reasons: we don't know there are how many type 2 strings. More important, those functions are unimportant in SAT problem.

As we know in intuition, it's very easy to distinguish the function outputs $2^n$ zeros (we call it reference function) with the function outputs $2^n$ ones. In the same time it's hard to distinguish the reference function with the function outputs $2^n - 1$ zeros and 1 one. The key problem here is how to measure the hardness (complexity) to distinguish the reference function and other different functions. We made a try to measure the complexity in the following way. Suppose we have two functions, one is the reference function, the other is the function outputs $2^n$ ones. They both can be represented by simple (**P** complexity long) SAT formula. In the beginning we cannot tell which is which. In order to distinguish them, what we need to do is to random choose one SAT formula, meanwhile random choose one input, then do the computation, the result will be either 0 or 1. In this way, we get 1 bit information. And this 1 bit information is enough to distinguish these two functions. We call that the function outputs $2^n$ ones has 1/2 bit information difference with reference function. Now we consider the general cases, suppose the function to be distinguished with reference function outputs $k$ ones and $2^n - k$ zeros. We still random choose one SAT formula, meanwhile random choose one input, then do the computation, there are totally $2^n + 2^n = 2^{n+1}$ possible input-outputs, in which $k$ outputs are one. To find the first output one in $2^n + 2^n = 2^{n+1}$ space, the information we get will be $n+1-k = log\frac{2^{n+1}}{k}$ bit in order to distinguish them. Or we can think like this: we have to do the computation $\frac{2^{n+1}}{k}$ times (the expected value) before we meet output 1, but before that the output are all zeros, and these zeros can be compressed to only $log\frac{2^{n+1}}{k}$ bit.

Now we can see the type 2 functions are unimportant in SAT problem because they are easy to be distinguished with reference function. While for the type 1 sequences, the situation is quite different because they contain all the hard functions to be distinguished with reference function. Then an important question arose, can type 1 sequences be generated by certain SAT formula?

If $k = O(P(n)),\ \ (k \ll 2^{n-1})$, the function of type 1 can be represented by a succinct SAT formula in disjunctive normal form (DNF). In order to compute a certain Boolean function $f(a)$, Consider its minterms $f^{(l)}(a)$ ,defined, for each such that $f(a^{(l)})$=1, as

$$f^{(l)}(a)=\begin{cases} 1 & if\ a = a^{(l)} \\ 0 & otherwise \end{cases}$$

Then the function $f(a)$ reads as follows:

$$f(a)=f^{(1)}(a) \vee f^{(2)}(a) \vee ... \vee f^{(k)}(a) \qquad (6)$$

Where $f(a)$ is the logical OR of all $k$ minterms. If $a^{(l)} = 110100...001$, we have

$$f^{(l)}(a)=a_{n-1} \wedge a_{n-2} \wedge \bar{a}_{n-3} \wedge a_{n-4} \wedge \bar{a}_{n-5} \wedge \bar{a}_{n-6} \wedge ... \wedge \bar{a}_2 \wedge \bar{a}_1 \wedge a_0$$

The length of this SAT formula (6) is approximate $O(kn)$.

Of course, such DNF formula is very easy to be distinguished in SAT problem as it's obviously satisfiable. But if we concatenate this DNF formula, using "AND" operator, with a formula generate the sequence of all "1", the composited formula will be not easy to be distinguished again. Anyway, in this way, we showed succinct formula does exist, which can generate type 1 sequence and in the same time cannot be distinguished directly or easily.

In the same time, we have the following program [4] which can generate the Type 1 strings (Program 2).

> Generate, in lexicographic order, all sequences with *k* ones; Of these sequences, print the *I* th sequence.

This program will print out the required string. The only variables in the program are *k* (with known range $\{0,1,...2^n\}$) and *I* (with conditional range $\{1,2,...,C_{2^n}^k\}$). The total length of this program is:

$$l(p)=c+log\,2^n+log\,C_{2^n}^k\le c'+log\,2^n+2^nH(\frac{k}{2^n})-\frac{1}{2}log\,2^n \tag{7}$$

The Kolmogorov complexity of this binary string x is bounded by [4]

$$K(x_1,x_2,...,x_{2^n}|2^n)\le 2^nH(\frac{k}{2^n})+\frac{1}{2}n+c \tag{8}$$

In which $H(p)=-p\log p-(1-p)\log(1-p)$. In this way, they can be compressed to at least $\frac{1}{2}n+2^nH(\frac{k}{2^n})$ bit according to Kolmogrov complexity theory. With regard to the length of Program 1 which can generate the same sequence, for succinct SAT formula of type 1, now we have another upper bound of the same string's Kolmogrov complexity. They should have the same level of complexity, especially when there are only two complexity classes, polynomial and exponential length**,** to be distinguished. We can get the following result:

$$O(P(n))=2^nH(\frac{k}{2^n})+\frac{1}{2}n+c\Rightarrow 2^nH_0(\frac{k}{2^n})=O(P(n)) \tag{9}$$

Unfortunately we cannot get the theory analysis of equation (9), we did a little experiment and just report the results that we think important. In Fig 1: the ordinate is $y=2^nH_0(\frac{k}{2^n})$ while abscissa is *n*. we can see clearly that $y=2^nH_0(\frac{k}{2^n})$ increase linearly when *k* is fixed and $k<<2^n$.

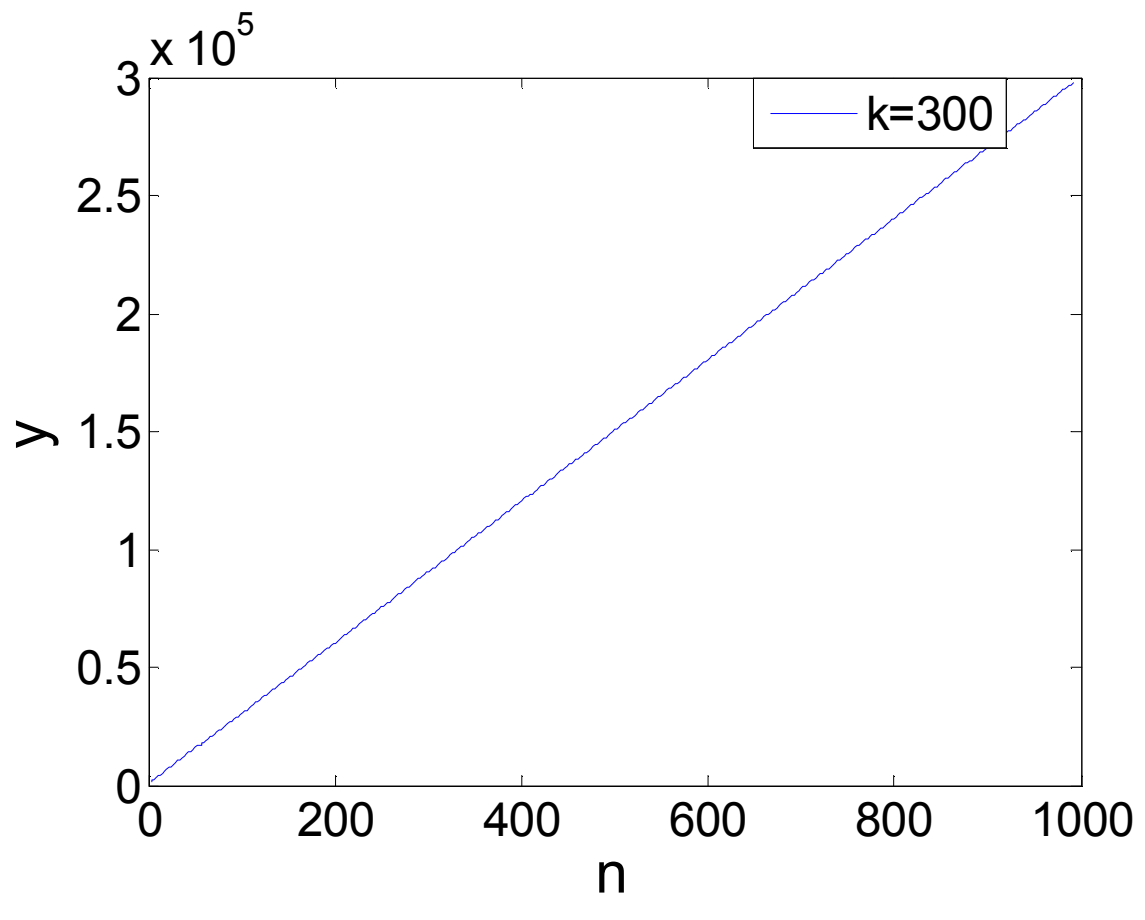

Figure 1. The property of equation (9)

## 4. THE BASIC PROBLEM AND COMPLEXITY OF SAT PROBLEM

Suppose that a computer is fed a random program. The probability distribution on the output strings is far from uniform[4]. Under the computer-induced distribution, simple strings are more(exponentially) likely than complicated strings of the same length. This motivated the definition of universal probability distribution on strings as follows:

The universal probability of a string $x$ is:

$$P_U(x) = \sum_{p:U(p)=x} 2^{-l(p)} = \Pr(U(p) = x) \tag{10}$$

which is the probability that a program randomly drawn as a sequence of fair coin flips $p_1, p_2, \ldots$ will print out the string $x$. This probability is universal in many senses. It can be considered as the probability of observing such a string in nature [4]; the implicit belief is that simpler strings are more likely than complicated strings.

Kolmogrov complexity and universal probability have equal status as universal algorithmic complexity measures. Their relation is very simple:

$$P_U(x) \approx 2^{-K(x)} \tag{11}$$

This is especially interesting since $log\dfrac{1}{P_U(x)}$ is the ideal codeword length (the Shannon codeword length) with respect to the universal probability distribution $P_U(x)$.

Suppose a computer is fed a succinct random program which can halts and in the same time output a $2^n$ long string $x$. As we already knew, this succinct random program must correspond to some succinct function or SAT formula. This string $x$ must can be greatly compressed and its Kolmogrov complexity $K(x) << 2^n$. The universal probability of this program (function, SAT formula) is $P_U(x) \approx 2^{-K(x)} >> 2^{-2^n}$.

Now we only consider the type 1 succinct string $X_k = \{x_1, x_2, \ldots, x_{2^n}\}$, and $\sum_{i=1}^{2^n} x_i = k$.

With equation (8), their Kolmogorov complexity is:

$$K(X_k) \le 2^n H(\frac{k}{2^n}) + \frac{1}{2}n + c \tag{12}$$

$$P_U(X_k) \approx 2^{-K(X_k)} \ge 2^{-2^n H(\frac{k}{2^n}) - \frac{1}{2}n - c} \tag{13}$$

In this paper we take those SAT formulas that generate sequences with same entropy as a set(ensemble). There are two reasons: their Kolmogorov complexities are same, their universal probability are also same, if we randomly generate SAT formula, they are generated with equal possibility. More important, according to Theorem 2, the sequences generated by them are Bernoulli distributions with same parameter. From the view of SAT formula ensemble we define the following basic problem.

**The basic problem:** Distinguish two different Bernoulli distributions that are generated by two SAT formula ensembles.

The SAT problem can be seen as composed by the basic problem. As unsatisfiable SAT formula ensemble correspond to $\text{Bernoulli}(0)$ distribution, while satisfiable SAT formula ensemble correspond to $\text{Bernoulli}(\gamma),(0<\gamma\le 1)$ distribution. To solve SAT problem is to distinguish the $\text{Bernoulli}(0)$ distribution with other $\text{Bernoulli}(\gamma),(0<\gamma\le 1)$ distributions, of course these distributions are all induced under computer with certain SAT ensemble. We are especially interested in those case when $\gamma$ is close to 0.

Then we can use the universal probability to define the complexity of SAT problem.

$$Complexity\ of\ SAT=\sum_k P_U(E_k).C(E_k),\ k=2^n.\gamma \qquad (14)$$

In which $E_k$ represent the SAT formula ensemble that generate the sequence with $k$ "1". $C(E_k)$ represent the computation complexity corresponded to distinguish $E_k$ ensemble with reference ensemble.

As we know in intuition, the computation complexity increase when $\gamma$ is close to 0. The most natural thought is to define certain distance in space. If we can define certain distance to measure the "closeness" of both distribution and SAT formula ensemble, the complexity might be connected to that distance. Here are two spaces in this problem. We call them SAT formula(program) space and probability(information) space. Before computation, we are in the SAT formula space, after computation we are in the probability space. These two spaces are transformed by the computation process. Then two most important questions naturally arose. How to define this distance and what is the distance related to the computation complexity.

It's quite interesting there is a quantum version of the basic problem, though not expressed in computation terms, has been well studied for over 20 years[5]. We'll postpone the SAT problem to the end and first discuss several relevant basic concept.

## 5. QUANTUM VERSION OF THE BASIC PROBLEM

Our original idea is to map the above basic problem into certain physics and wish to deduce some corresponded physical lower bound. It's very lucky this basic problem has already been thoroughly studied in the name statistical distance[5]. What we need to do is to directly use their result. Because the paper by W.K. Wootters is very easy to comprehend and directly connected to the basic problem. We'll mainly and directly use his result, even citing his original words in many place.

If one tosses a coin 100 times and finds that "heads" occurs 30 times, he will conclude that the probability of heads is roughly 0.30 (the coin is weighted unusually). However, because of the unavoidable statistical fluctuations associated with a finite sample, he cannot know the value of this probability exactly. In the above example the probability of heads may well be around 0.26 or 0.34. The same thing happens in quantum measurements. If a finite ensemble of identically prepared quantum systems is analyzed by some fixed measurement device, the observed frequencies of occurrence of the various outcomes typically differ somewhat from the actual probabilities. Because of this statistical error, one cannot necessarily distinguish (in a fixed number of trials) between two slightly different preparations of the same quantum systems. We can say that two preparations are indistinguishable in a given number of trials if the difference in the actual probabilities is smaller than the size of a typical fluctuation. Now we have

**The quantum version basic problem:** Distinguish two different Bernoulli distributions that are induced by two preparations of quantum system.

Imagine a beam of photons prepared by a polarizing filter and analyzed by a nicol prism. Let $\theta \in [0, \pi]$ be the angel by which the filter has been rotated around the axis of the beam. Each photon, when it encounters the nicol prism, has exactly two options: to pass straight through the prism(call this "yes" outcome) or to be deflected in a specific direction characteristic of the prism(the "no" outcome). By counting how many photons yield each of the two possible outcomes, an experimenter can learn something about the value of $\theta$ via the formula

$$p = cos^2 \theta \tag{15}$$

where $p$ is the probability of "yes".

Then, because of the statistical fluctuations associated with a finite sample ($m$), the observed frequency of occurrence of "yes" is only an approximation to the actual probability of "yes". More precisely, the experimenter's uncertainty in the value of $p$ is

$$\Delta p = \left[ \frac{p(1-p)}{m} \right]^{1/2} \tag{16}$$

This uncertainty causes the experimenter to be uncertain of the value of $\theta$ by an amount

$$\Delta\theta = \left| \frac{dp}{d\theta} \right|^{-1} \Delta p = \left| \frac{dp}{d\theta} \right|^{-1} \left[ \frac{p(1-p)}{m} \right]^{1/2} \tag{17}$$

Thus each value of $\theta$ can be associated with a region of uncertainty, extending from $\theta - \triangle\theta$ to $\theta + \triangle\theta$. Let us call two neighboring orientations $\theta$ and $\theta'$ distinguishable in $m$ trials if their regions of uncertainty do not overlap, that is, if

$$|\theta-\theta'|\geq\Delta\theta+\Delta\theta' \quad (18)$$

This idea of distinguishability is then used to define a notion of distance, called "statistical distance" between quantum preparation. The statistical distance $d(\theta_1,\theta_2)$ between any two orientations $\theta_1$ and $\theta_2$ is defined to be

$$d(\theta_1,\theta_2)=\lim_{m\to\infty}\frac{1}{\sqrt{m}}\times[\text{maximum number of intermediate orientations each of which is distinguishable (in } m \text{ trials) from its neighbors}] \quad (19)$$

This statistical distance is obtained essentially by counting the number of distinguishable orientations between $\theta_1$ and $\theta_2$.

From Eqs. (17)-(19) we obtain the following expression for statistical distance in terms of the function $p(\theta)$

$$d(\theta_1,\theta_2)=\frac{1}{\sqrt{m}}\int_{\theta_1}^{\theta_2}\frac{d\theta}{2\Delta\theta}=\int_{\theta_1}^{\theta_2}d\theta\frac{|dp/d\theta|}{2[p(1-p)]^{1/2}} \quad (20)$$

Upon substituting the actual form of the probability law $p(\theta)=cos^2\theta$ into this expression, the statistical distance become

$$d(\theta_1,\theta_2)=\theta_2-\theta_1; \quad (21)$$

In fact, the concept of statistical distance can be defined on any probability space and is quite independent of quantum mechanics. In a case where there are exactly two possible outcomes, the probability space is one-dimensional, every coin being characterized by its probability of heads. the statistical distance $d(p_1,p_2)$ between two coins with probability $p_1$ and $p_2$ of heads is defined in the following way:

$$d(p_1,p_2)=\lim_{n\to\infty}\frac{1}{\sqrt{m}}\times[\text{maximum number of mutually distinguishable (in } m \text{ trials) intermediate probabilities}] \quad (22)$$

Here two probabilities $p$ and $p'$ of heads are called distinguishable in *m* trials if

$$|p-p'|\geq\Delta p+\Delta p', \quad (23)$$

where, as before

$$\Delta p=\left[\frac{p(1-p)}{m}\right]^{1/2} \quad (24)$$

then the following expression for statistical distance can be obtained:

$$d(p_1, p_2) = \int_{p_1}^{p_2} \frac{dp}{2[p(1-p)]^{1/2}} = cos^{-1}(p_1^{1/2} p_2^{1/2} + q_1^{1/2} q_2^{1/2}) \tag{25}$$

where $q_1 = 1 - p_1$ and $q_2 = 1 - p_2$.

One can obtain the statistical distance by counting the number of these curves that "fit" between two given points in probability space. In a sense this is the most natural notion of distance on probability space, since it takes into account the actual difficulty of distinguish different probabilistic experiments. in the same time, the statistical distance between $p_1$ and $p_2$ is the shortest distance along the unit sphere. This shortest distance is equal to the angel between the unit vectors.

The main result of [5] is that the (absolute) statistical distance between two preparations is equal to the angel in Hilbert space between the corresponding rays.

**Theorem 4**: The statistical distance in Hilbert-space and probability space are equivalent and invariant under all unitary transformations .

The angel in Hilbert space is the only Riemannian metric on the set of rays, up to a constant factor, which is invariant under all unitary transformations (computed by any quantum computer), that is, under all possible time evolutions. In this sense, it is a natural metric on the set of states. It's as if nature defines distance between states by counting the number of distinguishable intermediate states.

As we know, a quantum computation is composed of three basic steps [6]: preparation of the input state, implementation of the desired unitary transformation acting on the state and measure of the output state.

If quantum computer is used to solve the quantum version basic problem, with Theorem 4 we can get an important conclusion. It's no difference to distinguish the two preparations in Hilbert space or probability space, because the angel in Hilbert space is invariant under all unitary transformations. Then how many measurement must be performed in order to solve it?

Suppose $p_1 = 0$, we have

$$\Delta p_1 = \left[\frac{p_1(1-p_1)}{m}\right]^{1/2} = 0; \qquad \Delta p_2 = \left[\frac{p_2(1-p_2)}{m}\right]^{1/2} \tag{26}$$

in order to distinguish them, we have

$$\begin{gathered} |p_2 - p_1| \geq \Delta p_1 + \Delta p_2 \Rightarrow \\ p_2 \geq \left[\frac{p_2(1-p_2)}{m}\right]^{1/2} \Rightarrow m \geq \frac{1-p_2}{p_2} \Rightarrow if\ p_2 = 2^{-n}\ then\ m \geq 2^n - 1 \end{gathered} \tag{27}$$

The expected trails in order to distinguish two preparations will increase exponentially with *n*. In the case of quantum computation, this means if two preparations are close enough, $|p_2 - p_1| = 2^{-n}$, the measurement must be performed exponential times in order to distinguish them. In this way, the quantum version basic problem cannot be efficiently solved by any quantum computer.

## 6. THE STATISTICAL DISTANCE IN SAT FORMULA SPACE

Now return to the SAT problem, there are also two spaces: the SAT formula space and probability space. There is definite mathematical connection between the ubiquitous statistical fluctuations in the outcome of computation and the set of SAT formulas.

Owing to the similarity with the quantum version basic problem, we guess the following theorem holds.

**Theorem 5:** The statistical distances in SAT formula space and probability space are identical and invariant under any computation.

In order to prove this theorem, we must first define a proper distance in SAT formula space. Kolmogorov complexity(universal probability) seems a perfect candidate at first sight, especially when universal probability itself corresponds to a probability space. But there is a serious flaw, the same Kolmogorov complexity could not distinguish Bernoulli $(\theta)$ and Bernoulli $(1\text{-}\theta)$ distribution. We still don't know the concrete statistical distance form in SAT formula space. We can only use the original definition of statistical distance.

We define the statistical distance $d\,(f_1, f_2)$ between two SAT formula ensembles as

$$d(f_1, f_2) = \lim_{m\to\infty}\frac{1}{\sqrt{m}} \times [\text{maximum number of intermediate SAT formula ensembles each of which is distinguishable (in } m \text{ trials) from its neighbors}] \tag{28}$$

And we have

$$\Delta f = \left|\frac{dp}{df}\right|^{-1} \triangle p = \left|\frac{dp}{df}\right|^{-1}\left[\frac{p(1-p)}{m}\right]^{1/2} \tag{29}$$

$$| f - f' | \geq \triangle f + \triangle f' \tag{30}$$

According to Theorem 2, since the mapping from SAT formula ensemble space to probability space is one to one. We'll have

$$\begin{aligned} d(f_1, f_2) &= \frac{1}{\sqrt{m}}\int_{f_1}^{f_2}\frac{df}{2\triangle f} = \int_{f_1}^{f_2} df\,\frac{|dp/df|}{2[p(1-p)]^{1/2}} \\ &= \int_{p_1}^{p_2}\frac{dp}{2[p(1-p)]^{1/2}} = d(p_1, p_2) \end{aligned} \tag{31}$$

Theorem 5 showed there is no difference to solve the basic problem in SAT formula space or probability space, as their statistical distances are identical. Same with quantum version basic problem, we can reach the same inference. At the worst case, any computer plus algorithm must need exponential trials to solve it. We can get the final conclusion: NP != P .

We made an comparison of the basic problem in Tab.1.

Table 1: The comparison of quantum and SAT formula version basic problem

| Basic problem | Quantum version | | SAT formula version | |
|---|---|---|---|---|
| space | Hilbert space | Probability space | SAT formula space | Probability space |
| distance | Angel | Statistical distance | ? | Statistical distance |
| Connection between two space | $p = cos^2 \theta$ | | Theorem 2 | |
| Equivalence of distance | Theorem 4 | | Theorem 5 | |

## 7. CONCLUSIONS

In this paper we have shown an interesting connection between Kolmogorov complexity and computational complexity. In our opinion, the equivalence of statistical distance in SAT formula space and probability space is the key to NPC problem.